\documentclass[article,authoryear,ipht]{beg_39}             

\usepackage[hang]{footmisc}
\usepackage{siunitx}
\fancypagestyle{plain}{%
  \fancyhf{}
  \fancyhead[R]{\small {\it \jname}, x(x): \thepage--\pageref{LastPage} (\myyear\today)}
  \fancyfoot[R]{\small\bf\thepage }
  \fancyfoot[L]{\fottitle}
  }
\renewcommand{\dmyy}{17}
\renewcommand{\myyear}{2023}
\renewcommand{\today}{}
\begin{document}

\volume{Volume x, Issue x, \myyear\today}
\title{Experimental Study of Condensation of Water on Polydimethylsiloxane-Coated Copper Surfaces}
\titlehead{Experimental Study of Condensation of Water on Polydimethylsiloxane-Coated Copper Surfaces}
\authorhead{T. Pfeiffer, S. Li, M. Kappl, H. Butt, P. Stephan \& T. Gambaryan-Roisman}
\corrauthor[1]{Till Pfeiffer}
\author[1]{Shuai Li}
\author[2]{Michael Kappl}
\author[2]{Hans-Jürgen Butt}
\author[2]{Peter Stephan}
\author[2]{Tatiana Gambaryan-Roisman}
\corremail{pfeiffer@ttd.tu-darmstadt.de}
\corraddress{Institute for Technical Thermodynamics, Technical University of Darmstadt, Darmstadt, Germany, 64287}
\address[1]{Institute for Technical Thermodynamics, Technical University of Darmstadt, Darmstadt, Germany, 64287}
\address[2]{Physics at Interfaces, Max-Planck-Institute for Polymer Research, Mainz, Germany, 55128}

\abstract{Modification of surfaces to enable dropwise condensation is a promising approach for achieving high condensation rates. In this work, we present an experimental study on condensation of water on copper surfaces coated with an ultrathin, $ \SI{5}{nm}-\SI{10}{nm} $ thick polydimethylsiloxane (PDMS) layer. This hydrophobic coating possesses a very low thermal resistance, which in combination with copper substrate enables achieving high condensation rates in heat transfer applications. The PDMS-coated copper substrates have been fabricated with a newly developed method, which involves turning, sanding, polishing, oxidation, and polymer coating steps.  The measured static contact angle was $\SI{110}{\degree} \pm \SI{1}{\degree} $, and a the contact angle hysteresis was $\SI{2}{\degree}$. The achieved very low hysteresis is advantageous for promoting dropwise condensation. The surface showed no ageing effects during 100 repetitions of advancing and receding contact angle (ARCA) measurements.
Condensation heat transfer on uncoated and PDMS-coated copper surfaces surfaces has been studied experimentally in a saturated water vapor atmosphere at $\SI{60}{\celsius}$. An enhancement factor for heat flux and heat transfer coefficient of up to 1.6 was found on PDMS-coated copper surfaces compared to uncoated surfaces, which decreased to 1.1 on second and third day of condensation operation.\\
Images of the condensation surface were recorded while conducting condensation experiments and post processed to evaluate drop departure diameter and frequency of drop sweeping events.\\
It has been shown that the behavior of the heat transfer coefficient correlates with the frequency of the sweeping events.
}

\keywords{Dropwise condensation; Heat transfer; Surface fabrication}

\maketitle

\section{Introduction}
\label{sec:Introduction}
Water condensation is critical in various applications, such as atmospheric fresh water harvesting \cite{Bhushan.2020,Hou.2018,Jarimi.2020,Khalil.2016}, HVAC applications \cite{Cavallini.2003} and power conversion \cite{Beer.2007,Pan.2018,Zhao.1994}, where high condensation rates are essential. Dropwise condensation (DWC) is characterized by significantly higher condensation rates and higher heat transfer coefficients than filmwise condensation (FWC) \cite{Schmidt.1930}, since the continuous condensate film forming during FWC, due to its high thermal resistance, isolates the cool substrate from the vapor. In contrast, the thermal resistance of drops is lower and decreases with decreasing drop size. In the case of condensation on vertical or inclined surfaces, the drops roll off by gravity as they reach a sufficient size, known as drop departure diameter, providing free space for nucleation of new drops. The efficiency of DWC obviously increases with decreasing drop departure diameter and the related increase of the frequency of drop sweeping events.\\
The common materials used in condensers, such as copper, possess high thermal conductivity. These materials are usually hydrophilic and therefore promote the filmwise condensation. One promising approach to enhance the condensation rate is modifying the surface to achieve hydrophobic behavior with low contact angle hysteresis and thereby to facilitate DWC. However, the available hydrophobic coatings have shortcomings, either being thermally insulating, non-durable or both \cite{Ahlers.2019}.\\
Numerous studies on surface modifications to enable DWC have been conducted and different approaches with corresponding advantages and disadvantages have been reported in literature \cite{Attinger.2014,ElFil.2020,Goswami.2021,Hu.2021,Xu.2021}. The most commonly investigated surface coating techniques to enable DWC are polymer coatings, self-assembled monolayers (SAMs), ion-implantation and lubricant infused surfaces (LIS).\\
Polymer coatings have already been thoroughly studied \cite{ORKANUCAR.2013}. They combine advantages of hydrophobic wetting behavior with high coating durability \cite{Holden.1987} and well established fabrication methods \cite{Goswami.2021}. Common polymers used to coat surfaces in condensation heat transfer applications are PTFE \cite{Ma.2002}, PFDA \cite{Paxson.2014} and polymer composites \cite{Damle.2015}. Nevertheless, since the thermal conductivity of common polymers is low, the polymer coatings constitute a thermal resistance between a subcooled wall and vapor phase. Therefore, a trade-off between the stability of coating and heat transfer enhancement must be found when choosing the thickness of the polymer coatings \cite{Holden.1987}. \\
Self-assembled monolayers (SAMs) consist of a monomolecular layer of organic polar chemical compound, which adsorbes to a metallic substrate and shows hydrophobic wetting behavior. Due to their thickness in the range of tens of angstroms, SAMs thermal resistance is low compared to that of coatings fabricated by other techniques \cite{Das.2000,Yang.2006}. Nevertheless, the low coating thickness makes SAMs susceptiple to coating defects, ageing effects and with that strong decline of condensation heat transfer enhancement in a matter of hours of operation \cite{Vemuri.2006}.\\
For ion-implanted surfaces, the promotion of dropwise condensation takes place without deterioration of thermal contact between the subcooled substrate surface and the vapor. During ion-implantation, foreign element ions are inserted into metal surfaces and, by that, reduce the surface energy leading to a change in wetting behavior from hydrophilic to hydrophobic. Studies conducted on various substrate materials with different foreign element ions prove the versatility of this method, yet its high costs make the process economically unattractive \cite{Leipertz.2008,Qi.1991}.\\
Lubricant-infused surfaces (LISs) consist of a surface microstructure, which is flooded with a lubricant in order to promote drop mobility and thereby facilitate DWC. Current works aim at minimizing the lubricant's thermal resistance and continuous draining during operation \cite{Tripathy.2021,Weisensee.2017}.\\
One polymer coating, which has recently gained attention in DWC research, is trimethylsiloxy terminated linear (poly)dimethylsiloxane, PDMS \cite{Eduok.2017,Zhang.2020}. In addition to hydrophobicity, PDMS has shown a potential for self healing \cite{Xue.2016}, which has been applied to fabrics to maintain self-cleaning properties and water repellency over multiple washing and abrasion cycles. Teisala et al. \cite{Teisala.2020} developed ultrathin PDMS brush coatings, which combine multiple advantageous characteristics for condensation heat transfer applications. The low coating layer thickness of several nanometers ensures a significantly reduced thermal resistance compared to other polymer coatings. The low thermal resistance of PDMS brushes and favorable wetting properties (hydrophobicity and low contact angle hysteresis), resulting in DWC mode promise great potential to enhance condensation heat transfer. The coating procedure has been developed for glass slides and polished silicon wafers as substrates, taking advantage of the fact that the methyl-terminated polydimethylsiloxane spontaneously bonds on materials with silicon oxide surface chemistry at moderate temperatures. These types of substrates, however, have only limited practical applications \cite{LO20192806}.\\
Fazle Rabbi et al. \cite{FazleRabbi.2022} investigated PDMS to promote DWC on copper, aluminum and stainless steel during ethanol, hexane and pentane condensation.
In order to enable DWC and prevent condensate penetration of the PDMS coating, a coupling of flexible PDMS with low energy silane was investigated. The developed coating was applied to the aforementioned metal substrates, and condensation heat transfer enhancement of 274\% for ethanol, 347\% for hexane and 636\% for pentane compared to uncoated surfaces was observed on coated horizontal tubes. The DWC mode was sustained for 15 days of steady operation without signs of coating degradation. 
While in this study the results of detailed condensation heat transfer measurements for an extended experimental time scale are reported, no quantitative characteristics of wetting behavior during the condensation experiment (such as the drop departure diameter or frequency of sweeping events) have been determined in the study. \\
Li et al. \cite{Li.2022} investigated wetting and heat transfer during condensation of water-ethanol mixtures of different compositions on PDMS-coated substrates. By introducing a small water fraction of 10\% into the vapor phase, an increase in heat transfer coefficient of 1800\% from $\SI{3}{kW \per m^2 K}$ for pure ethanol to $\SI{57}{kW \per m^2 K}$ for the mixture was observed. In addition, the coalescence velocity for neighboring condensate droplets increased by 170\% by introducing the aforementioned water fraction, which led to a quicker re-exposure of the subcooled wall to the vapor phase and thus to a heat transfer enhancement.
While this study lead to deeper insight into coupled surface wetting and heat transfer enhancement, the effects responsible for the higher condensation efficiency lay in the mixture-nature of the water ethanol composition. No investigations on pure water as condensing fluid have been conducted in that framework.\\
In spite of the increasing interest in the scientific community for heat and mass transfer during dropwise condensation, the simultaneous quantitative investigations of heat transfer characteristics and the characteristics of drop dynamics during condensation, such as drop coalescence, drop departure diameter and frequency of drop sweeping events, are very rare \cite{Li.2022, Tripathy.2021}. While Rose \cite{Rose.1976} investigated surface wetting behavior during DWC, characterized by sweeping frequency, maximum drop size and nucleation site density, the heat transfer data is only measured for varying maximum drop sizes. No statement on the relation to heat transfer coefficient is made for the other wetting characteristics.
Merte et al. \cite{Merte.1983} investigated drop profile, defined by advancing and receding contact angle, and departure diameter during DWC, but no quantitative deductions on the heat transfer coefficient could be made.
The lack of those combined data limits the understanding of mechanisms determining the condensation rates and, therefore, the methods for optimization of condensation efficiency.  
In this work, condensation of water vapor on copper surfaces coated with ultrathin PDMS layers is studied experimentally. The fabrication method of PDMS layers previously developed for glass and silicon substrates \cite{Teisala.2020} has been extended and adapted to copper surfaces which are commonly used in industrial applications. The heat transfer characteristics during condensation are measured simultaneously with capturing the dynamics of condensate on the substrate. This allows establishing a direct relationship between the measured heat flux and the heat transfer coefficient, on one side, and the wetting behaviour and dynamics of drops on the other side. The heat transfer enhancement in comparison to uncoated copper surfaces is reported. 
The ageing effects of PDMS-coated copper substrates during the condensation experiment, which are of fundamental interest for heat exchanger applications and their use in industry, are investigated by comparison of contact angles before and after the condensation operation, as well as by following the temporal evolution of characteristics of drop dynamics, such as the drop departure dynamics and frequency of the sweeping events, during the condensation.
\section{Experimental methods}
\label{sec:ExpMethods}
\subsection{Surface fabrication and characterization}
\label{sec:surfacefabrication}
\begin{figure}[!ht]
	\begin{center}
		\includegraphics[width=0.95\textwidth]{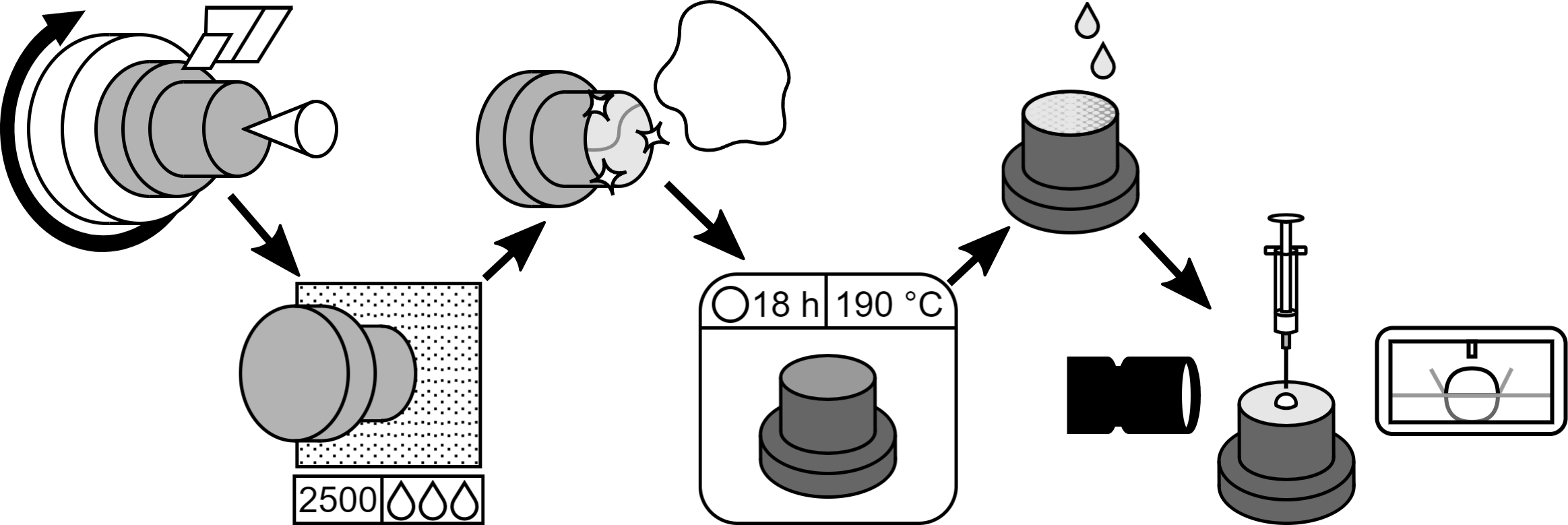}
		\put(-420,65){(1)}
		\put(-340,80){(2)}
		\put(-280,70){(3)}
		\put(-220,80){(4)}
		\put(-145,70){(5)}
		\put(-50,75){(6)}
		\caption{Schematic steps of the copper substrate fabrication and coating procedure}
		\label{FabricationProcedure}
	\end{center}
\end{figure}
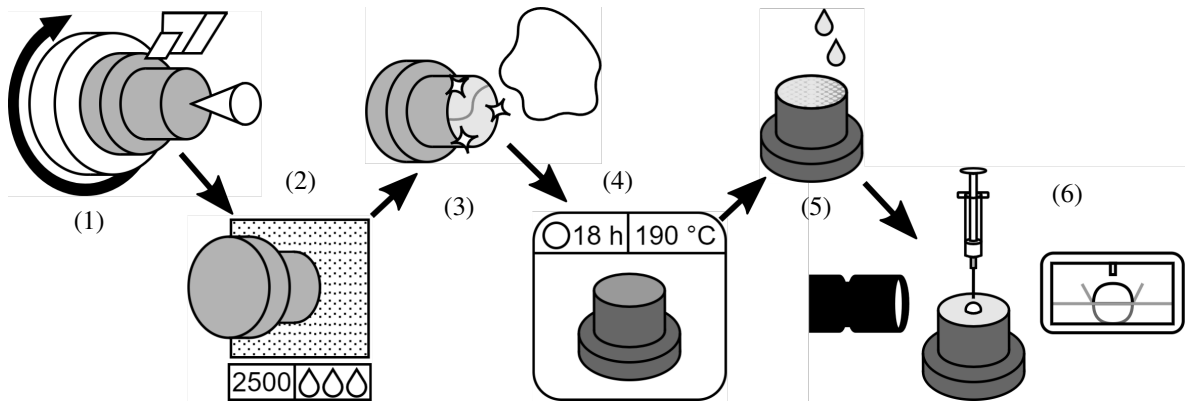

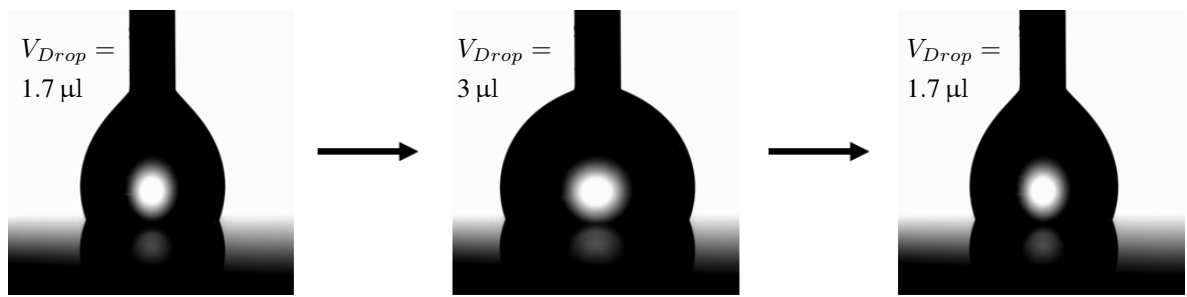
\begin{figure}[!ht]
	\begin{center}
		\includegraphics[width=0.95\textwidth]{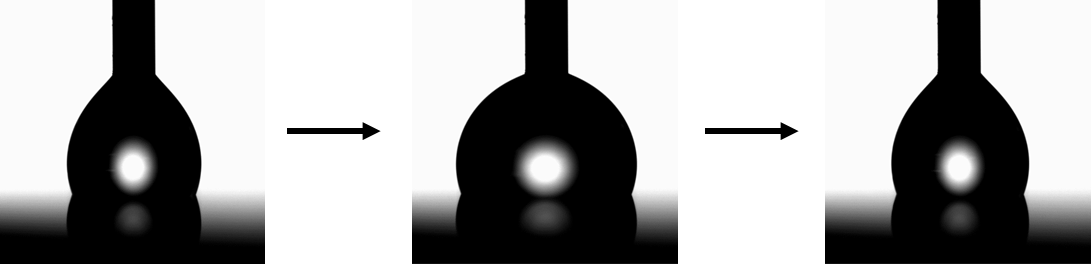}
		\put(-440,90){$ V_{Drop} = $}
		\put(-440,75){$\SI{1.7}{\mu l}$}
		\put(-275,90){$V_{Drop} = $}
		\put(-275,75){$\SI{3}{\mu l}$}
		\put(-105,90){$V_{Drop} =$}
		\put(-105,75){$\SI{1.7}{\mu l}$}
		\caption{Drop expansion and reduction sequence during ARCA investigation}
		\label{ARCADrop}
	\end{center}
\end{figure}

The PDMS-coating protocol involves surface preparation, coating, and characterization. These steps are depicted in Figure \ref{FabricationProcedure}.
Initially, the copper body is turned from the bar stock (1). The condensation surface is then wet-sanded with increasing grit ranging from 400 to 2500 (2), until a surface roughness value of $ S_{a} = \SI{31}{nm} $ is achieved, where $ S_{a} $ represents the arithmetic mean height of the surface. The roughness values are assessed using a 3D confocal microscope ($\mu$surf expert, NanoFocus AG). Subsequently, polishing paste is applied to attain a mirror-like finish (3), and confocal microscopy analysis shows a surface roughness value of $ S_{a} = \SI{11}{nm} $ after the polishing process. The copper sample is then cleaned in an ultrasonic acetone bath for 10 minutes to eliminate organic residues and flushed with DI water afterwards. The substrate is subsequently oxidized in a ventilated muffle furnace at $\SI{190}{\celsius}$ for 18 hours to generate a uniform copper oxide layer on the condensation surface (4). The substrate is again placed in an ultrasonic acetone bath for 10 minutes following the oxidation and cleaning process, and a slight increase in surface roughness to $ S_{a} = \SI{22}{nm} $ is observed.\\
To coat the surface with PDMS, the copper surface is first cleaned with oxygen plasma (Diener Electronic Femto, $\SI{120}{W}$). Subsequently a drop of silicone oil (Sigma-Aldrich, $M=\SI{11740}{g/mol}$) is allowed to spread and left for $\SI{24}{h}$ at $\SI{100}{\celsius}$ (5). Afterwards, excess PDMS is removed from the surface by sonication of the sample in toluene, ethanol and water for 10 minutes in each liquid, as described by Teisala et al. \cite{Teisala.2020}, resulting in a $\SI{5}{nm}-\SI{10}{nm} $ thick hydrophobic coating.\\
The wetting behavior of the PDMS-coated copper surface is evaluated by measuring the contact angle using a dataphysics® OCA 25 optical contact angle measurement device (6). For the measurement of advancing and receding contact angles (ARCA), a sessile water droplet is deposited on the surface and its volume is increased from $\SI{1.7}{\mu l}$ to $\SI{3}{\mu l}$ and then decreased back to its initial value. The variation in droplet volume is carried out using a cannula that penetrates the drop cap. The procedure is displayed in Figure \ref{ARCADrop}. The advancing contact angle is determined as the three-phase contact line advances during the increasing drop volume phase, while the receding contact angle is determined as the three-phase contact line recedes during the decreasing drop volume phase.
\subsection{Setup and procedure for the investigation of the condensation process}
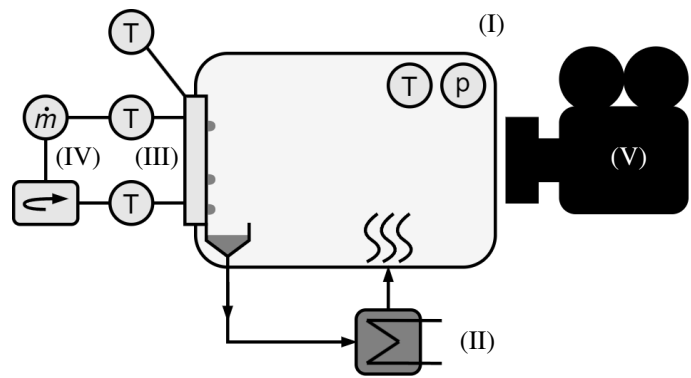
\begin{figure}[!ht]
	\begin{center}
		\includegraphics[width=90mm]{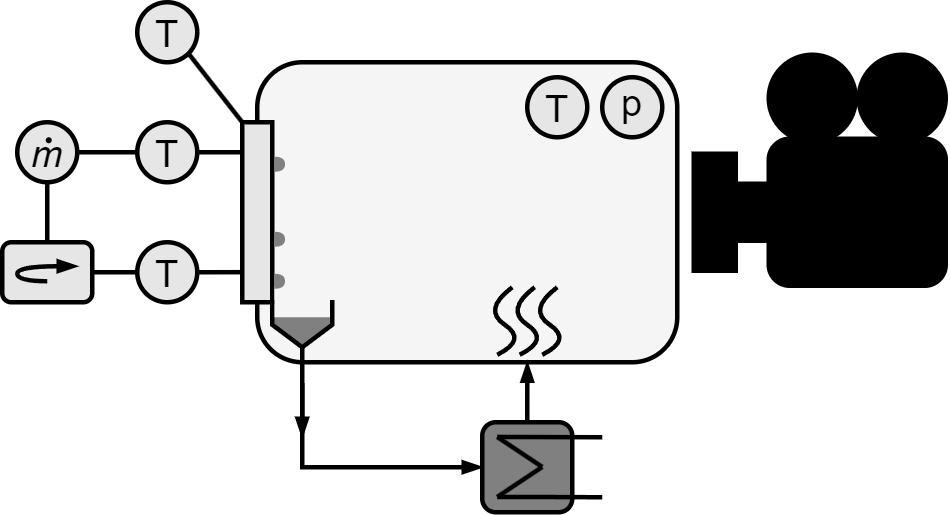}
		\put(-80,130){(I)}
		\put(-240,80){(IV)}
		\put(-210,80){(III)}
		\put(-87,10){(II)}\color{white}
		\put(-30,80){(V)}\color{black}
		\caption{Working scheme of the experimental setup}
		\label{WorkingScheme}
	\end{center}
\end{figure}

\begin{figure}[htbp]
		\begin{minipage}[t]{0.4\textwidth}
		\includegraphics[width=0.82\textwidth]{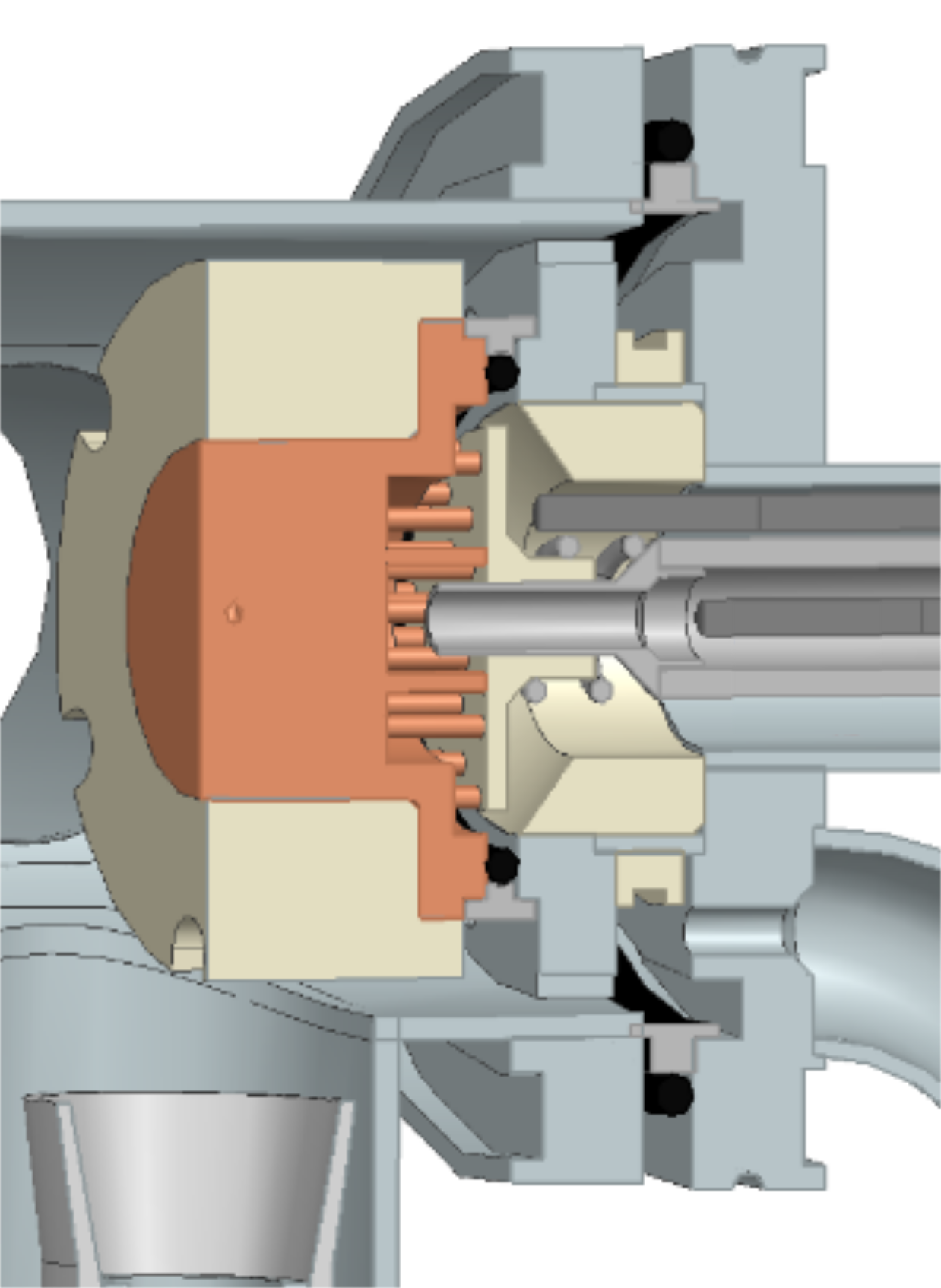}
		\color{white}
		\put(-76,106){(1)}
		\put(-104,98){(2)}
		\put(-32,106){(3)}
		\put(-32,124){(4)}\color{black}
		\caption{Cut view of the wall subcooling subsystem}
		\label{SubcoolingCut}
	\end{minipage}
	\begin{minipage}[t]{0.59\textwidth}
	\hspace{1.2cm}
		\includegraphics[width=0.8\textwidth]{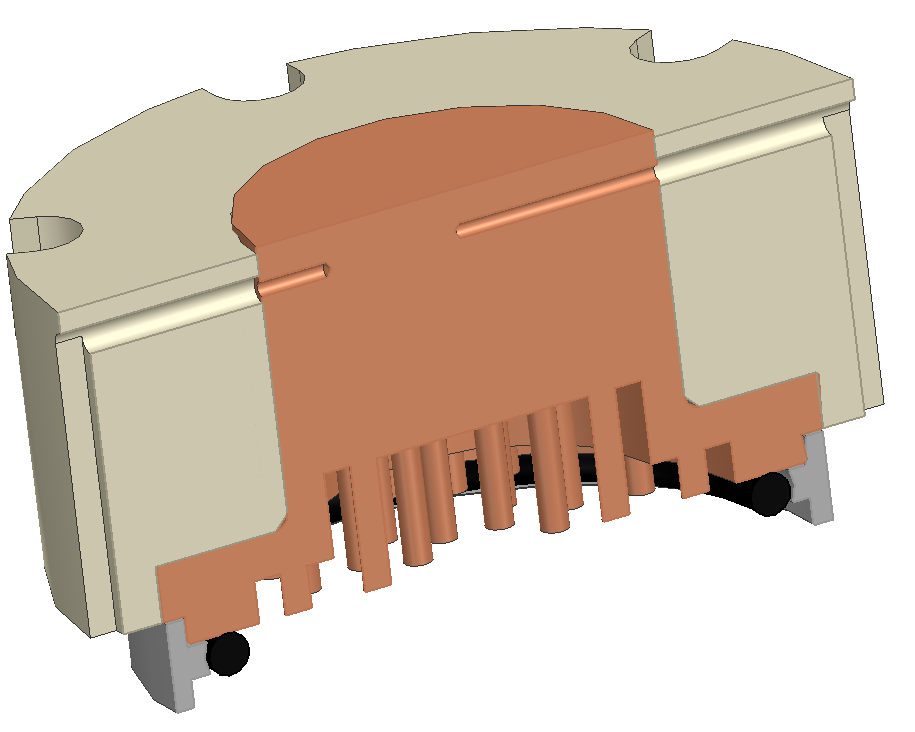}
		\put(-40,40){(A)}
		\put(-108,38){(B)}
		\put(-125,132){(C)}
		\put(-38,110){(D)}
		\put(-130,110){(E)}
		\caption{Cut view of the condensation substrate CAD model}
		\label{SubstrateCut}
	\end{minipage}
\end{figure}

The experimental setup used for this study is schematically illustrated in Figure \ref{WorkingScheme}.
The test cell (I) is custom-made from vacuum tubing parts and filled with water vapor at saturation conditions. The saturation conditions are continuously monitored using temperature and pressure sensors and maintained at a steady state by an externally controlled evaporator (II). The condensate generated from the hydrophobic, PDMS-coated copper substrate (III) under investigation is re-evaporated by the heater (II).
The PDMS-coated substrate (III) is sealed tightly on the subcooling subsystem with a coolant cycle (IV), including a Coriolis mass flow sensor (TMU S008, Heinrichs GmbH) and a chiller (Ministat 230, Huber GmbH), attached to lower the wall temperature and promote condensation. The coated condensation surface is monitored visually by a black and white camera (V) to record and analyze the wetting behavior. The data is recorded with an Andor Zyla 5.5 black and white camera with a modular Navitar Zoom 6000® lens system.\\
The wall subcooling subsystem, as shown in Figure \ref{SubcoolingCut}, comprises a central coolant feed line (1), which directs the coolant liquid to the backside of the coated substrate, where cooling fins (2) are located. The temperature of the coolant feed line is measured using a PT100 temperature sensor (3), while another PT100 temperature sensor (both $\SI{3}{mm}$ diameter, 1/10 Class B, THM GmbH) is used to measure the temperature increase of the coolant caused by the latent heat of condensation, as it passes through the coolant return line (4).
To investigate the wetting behavior and heat transfer performance of coated substrates during the condensation process, the copper substrate shown in Figure \ref{SubstrateCut} is utilized. The substrate has a plane face equipped with an O-ring and a centering ring (A), which effectively seals the coolant cycle against the pure water vapor atmosphere at saturation conditions inside the test cell. The substrate is designed with cooling fins (2 in Figure \ref{SubcoolingCut}, B in Figure \ref{SubstrateCut}) on the backside, which enhance the single phase heat transfer between the coolant and the copper substrate. The substrate features a visually accessible PDMS-coated condensation surface (C). Parasitic heat fluxes through surfaces other than the PDMS-coated condensation surface are minimized by press fitting the copper substrate into a PEEK insulation (D). The substrate has holes for four thermocouple type T sensors ($\SI{1}{mm}$ diameter, Type T, Class 1, THM GmbH), with two of them labeled (E) and the other two placed outside of the cut plane.\\
The following procedure is applied during the condensation experiments: First, the coated substrate is fixed on the coolant flange as shown in Figure \ref{SubcoolingCut}, tightly sealing the coolant cycle against the atmosphere inside the test cell. Next, the vacuum pump is connected to the test cell and the attached tubing to remove non-condensable gases (NCGs). The evaporator (II in Figure \ref{WorkingScheme}) is used to generate pure saturated water vapor, which fills the inside of the test cell and connected tubing. Subsequently, the coolant liquid temperature is reduced using a thermostat. The wall subcooling is stepwise increased to examine the heat transport at different levels of subcooling. In each step, when steady state is reached, 40 measurement values are accumulated and averaged. These averaged values are subsequently used to compare the condensation performance for different wall subcoolings and substrates.\\
Saturation conditions of $\SI{60}{\celsius}$ and $\SI{199}{mbar}$ were selected for the test campaign.
The saturation conditions remain constant throughout the experiment, while the coolant liquid temperature on the back side of the substrate is gradually reduced in increments of $\SI{7.5}{K}$. The accumulation of measurement values to be averaged starts 5 minutes after reaching steady state. No significant temporal changes take place during the period of data accumulation.\\
The coolant mass flow rate $\dot{m}_{cool}$ is measured using a Coriolis mass flow sensor. The inlet and outlet coolant temperatures are denoted as $T_{in}$ and $T_{out}$, respectively, and are measured at locations (3) and (4) as indicated in Figure \ref{SubcoolingCut} by PT100 sensors ($\SI{3}{mm}$ diameter, uncertainty 1/10 Class B, THM GmbH). The wall temperature, $ T_{wall}$, is determined by taking an average of readings from three thermocouples ($\SI{1}{mm}$ diameter, Type T, TMH GmbH) located at positions $\SI{3}{mm}$ below the condensation surface, as indicated by (E) in Figure \ref{SubstrateCut}. The vapor temperature, $ T_{vap}$, is measured using a PT100 sensor ($\SI{1.5}{mm}$ diameter, Class A, THM GmbH) located approximately $\SI{50}{mm}$ away from the condenser surface. All temperature sensors are calibrated to a reference with a measurement uncertainty of $\SI{30}{mK}$. The measurement frequency for all sensor signals is $\SI{0.36}{Hz}$.\\
The heat flux is determined through an energy balance for the coolant liquid on the back side of the substrate:
\begin{equation}
	\dot{q}=\dot{m}_{cool}c_{cool}(T_{in}-T_{out})/A_{cond} .
	\label{caloric}
\end{equation}
In equation (\ref{caloric}), the condensation surface area of the PDMS-coated copper substrate is represented by $ A_{cond} $, and $c_{cool}$ denotes the heat capacity of the coolant liquid (Lauda Kryo 51), which is determined from the data sheet as a function of temperature.
The heat transfer coefficient, $ \alpha $, is calculated according to the following expression:
\begin{equation}
	\alpha = \dot{q}/(T_{vap}-T_{wall}) ,
	\label{htc}
\end{equation}
where the heat flux $ \dot{q}$ is determined using equation (\ref{caloric}).\\
In addition, images of the vertical condensation surface were captured and analyzed to examine the surface wetting behavior during condensation. For this purpose, a series of 80 pictures of the entire substrate is taken with a recording frequency of $\SI{36}{Hz}$ at each subcooling step, with a spatial resolution of 678 pixels per centimeter, and then processed to determine the droplet departure diameter and frequency of the roll off and sweeping events.\\
The drop departure diameter is a crucial parameter for assessing the mobility of droplets and the ability of the surface to sustain DWC and renew nucleation sites. The departure diameter is determined by measuring the diameter of a drop about to be accelerated downwards due to gravity. This means that the recorded image sequence is evaluated with regard to departing droplets. A droplet is considered to be a departing droplet if it fulfills the following criteria: First, as soon as a droplet starts to move downward, it does not stop again due to reaching a pinning site or surface defect. Second, the departure event is initiated by apparent continuous growth and not by coalescence with a neighboring droplet of similar or comparable size or sweeping by a passing droplet from the top. To obtain a representative value for each subcooling step, three different departing drops are measured and the diameters are averaged.\\
Furthermore, the frequency of sweeping events in the central rectangular region of the substrate for each subcooling step is examined. It should be noted that the frequency of sweeping events is highly dependent on the location of the field of view on the substrate, since the presence of multiple droplets above a growing drop increases the likelihood of the drop being located in a sweeping path and therefore being removed from the surface. Here, a region of interest of $\SI{8}{mm}\times \SI{9.5}{mm}$ in the substrate center is evaluated. Each time a fully visible droplet rolls through the frame, the motion is registered as a sweeping event. Taking the duration of the recording and area into account, the frequency of sweeping events is then normalized in space and time.\\
In order to comprehensively assess changes in wetting behavior originating from the condensation operation on the PDMS-coated copper surface, static, advancing and receding contact angles, as well as the roll off angle are measured, both, before and after consecutive experimental runs. The static contact angle is measured by depositing a $\SI{2}{\mu l}$ water drop on the substrate. To evaluate the advancing and receding contact angles, as well as the roll off angle, a $\SI{30}{\mu l}$ drop is deposited on the surface, and the surface is gradually tilted starting from the horizontal state . Contrary to the presented ARCA procedure used to characterize the wetting behavior on freshly PDMS-coated substrates in Figure \ref{ARCADrop}, this procedure reliably quantifies surface wetting characteristics even on strongly aged surfaces, defined by high contact angle hysteresis and contact line pinning. When investigating these characteristics on the strongly aged surfaces with the aforementioned ARCA-procedure, no significant measurement values could be obtained due to strong contact line pinning. The aforementioned parameters allow insights in drop mobility on the PDMS-coated surface \cite{Extrand.1990,NanGao.}.
\section{Results and Discussion}
\label{sec:ResultsandDiscussion}
\subsection{Wetting of sessile droplets on PDMS-coated copper substrates}
\begin{figure}[htbp]
	\begin{minipage}[t]{0.37\textwidth}
		\includegraphics[width=0.95\textwidth]{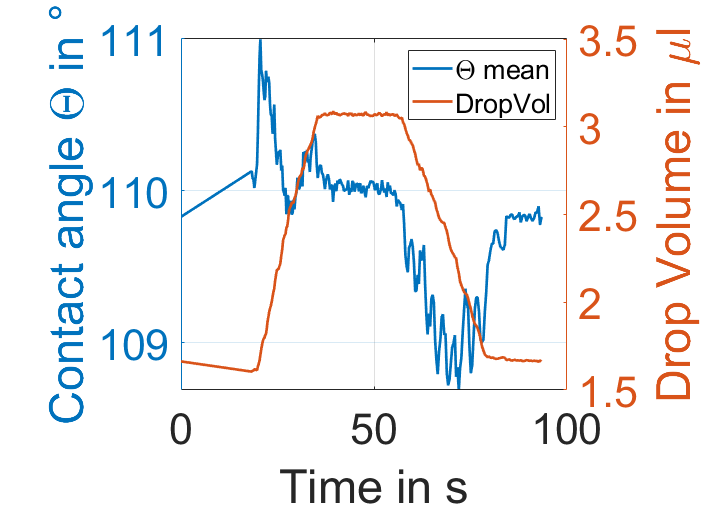}
		\caption{Contact angle and drop volume in single ARCA investigation}
		\label{1xARCA}
	\end{minipage}
	\hspace{1cm}
	\begin{minipage}[t]{0.62\textwidth}
		\includegraphics[width=0.79\textwidth]{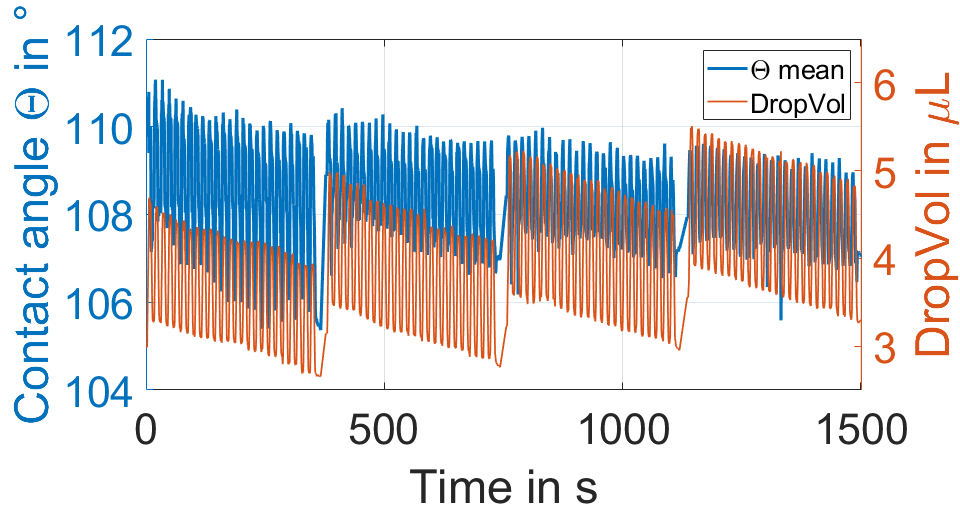}
		\caption{Contact angle and drop volume in 100x repeated ARCA\\ investigation}
		\label{100xARCA}
	\end{minipage}
\end{figure}

The results of the advancing and receding contact angle (ARCA) investigations on freshly prepared PDMS-coated copper substrates are displayed in Figures \ref{1xARCA} and \ref{100xARCA}.
Both figures depict the behavior of contact angle and drop volume during ARCA investigation. Specifically, Figure \ref{1xARCA} illustrates the relationship between the contact angle and the drop volume over time for a single ARCA measurement. During the drop expansion phase, as the drop volume increases from $\SI{1.7}{\mu l}$ to $\SI{3}{\mu l}$, the contact angle value rises to a peak of $\SI{111}{\degree} \pm \SI{1}{\degree} $. Subsequently, as the drop volume remains constant at $\SI{3}{\mu l}$, the contact angle value stabilizes at $\SI{110}{\degree} \pm \SI{1}{\degree} $. During the receding phase, as the drop volume decreases back to its initial value of $\SI{1.7}{\mu l}$, the contact angle value reduces to $\SI{109}{\degree} \pm \SI{1}{\degree} $, resulting in a contact angle hysteresis of $\SI{2}{\degree}$ for water on PDMS-coated copper surfaces.  These low hysteresis values, which are comparable with those measured on PDMS brush coatings on glass and silicon wafers \cite{Teisala.2020}  are very promising for condensation applications.
Figure \ref{100xARCA} shows the behavior of contact angle and drop volume over 100 repeated ARCA cycles. As drop volumes smaller than $\SI{1.5}{\mu l}$ tend to cause significant drop shape distortion due to the piercing canula, resulting in apparent contact angle deviations, the initial drop volume for repeated ARCA measurements is increased to $\SI{3}{\mu l}$ and periodically refilled after every 25 repetitions. This approach ensures that a sufficient drop volume is maintained throughout the measurement, counteracting drop evaporation that leads to a drop volume decrease, thus ensuring reliable and accurate results regarding the wetting behavior of water droplets on PDMS-coated copper surfaces.\\
During the first expansion phase, the drop volume is increased by $\SI{1.5}{\mu l}$ to a maximum value of $\SI{4.5}{\mu l}$ and an advancing contact angle of $\SI{111}{\degree} \pm \SI{1}{\degree} $ is recorded. During the first reduction phase, a receding contact angle of $\SI{108}{\degree} \pm \SI{1}{\degree} $ is recorded. Until the first water drop refill after 25 repetitions, the advancing contact angle value decreases to $\SI{110}{\degree} \pm \SI{1}{\degree} $ and the receding contact angle value decreases to $\SI{106}{\degree} \pm \SI{1}{\degree} $, resulting in a contact angle hysteresis of $\SI{4}{\degree}$.\\
During the three subsequent measurement blocks of 25 ARCA repetitions, the advancing contact angle value reduces from $\SI{110}{\degree} \pm \SI{1}{\degree} $ to $\SI{109}{\degree} \pm \SI{1}{\degree} $, while the receding contact angle value stays around $\SI{108}{\degree} \pm \SI{1}{\degree} $ to $\SI{107}{\degree} \pm \SI{1}{\degree} $, resulting in a contact angle hysteresis of $\SI{2}{\degree}$. The value obtained for the contact angle hysteresis after 100 cycles of passing three phase contact lines is close to the hysteresis value obtained from the single ARCA measurement, thereby confirming the durability of the PDMS-coating on the copper surface. These results are consistent with the wetting characteristics of PDMS-coated glass substrates, as reported by Teisala et al. \cite{Teisala.2020}, who observed advancing contact angles of $\SI{107}{\degree} \pm \SI{2}{\degree} $, receding contact angles of $\SI{105}{\degree} \pm \SI{2}{\degree} $, and a contact angle hysteresis of $\SI{2}{\degree} $. These findings provide further evidence for the effectiveness of PDMS-coatings in controlling surface wetting behavior.
\subsection{Condensate wetting on PDMS-coated copper substrates and ageing effects}
\begin{figure}[htbp]
 	\begin{minipage}[t]{0.49\textwidth}
 		\includegraphics[width=1\textwidth]{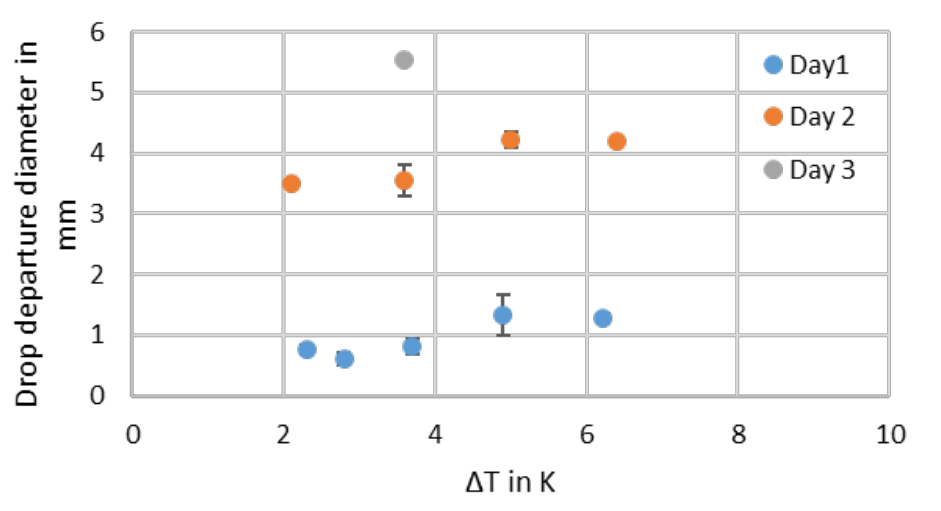}
 		\caption{Drop departure diameter over wall subcooling}
 		\label{DropDiaDep}
 	\end{minipage}
 	\begin{minipage}[t]{0.49\textwidth}
 		\includegraphics[width=1\textwidth]{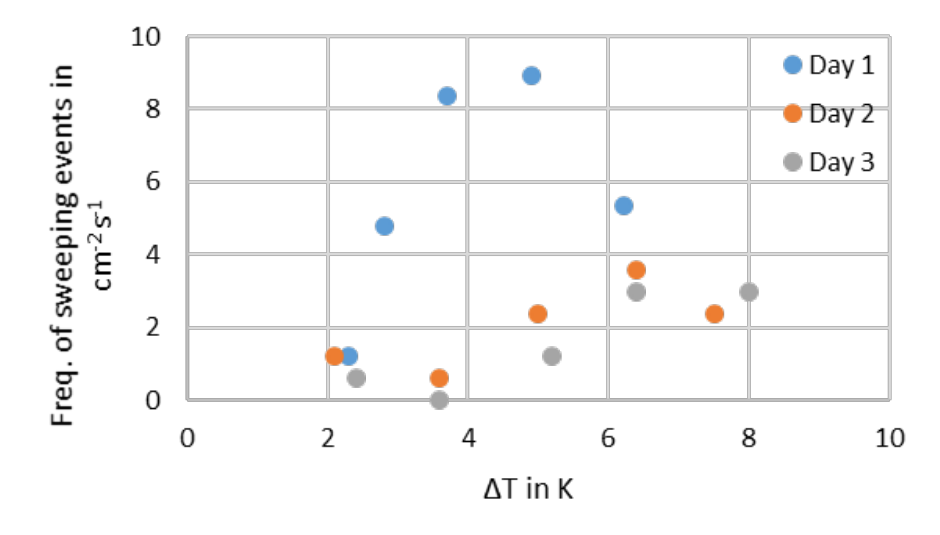}
 		\caption{Frequency of sweeping events over wall subcooling}
 		\label{FreqSweeping}
 	\end{minipage}
\end{figure}

\begin{figure}[!ht]
	\begin{center}
		\includegraphics[width=0.38\textwidth]{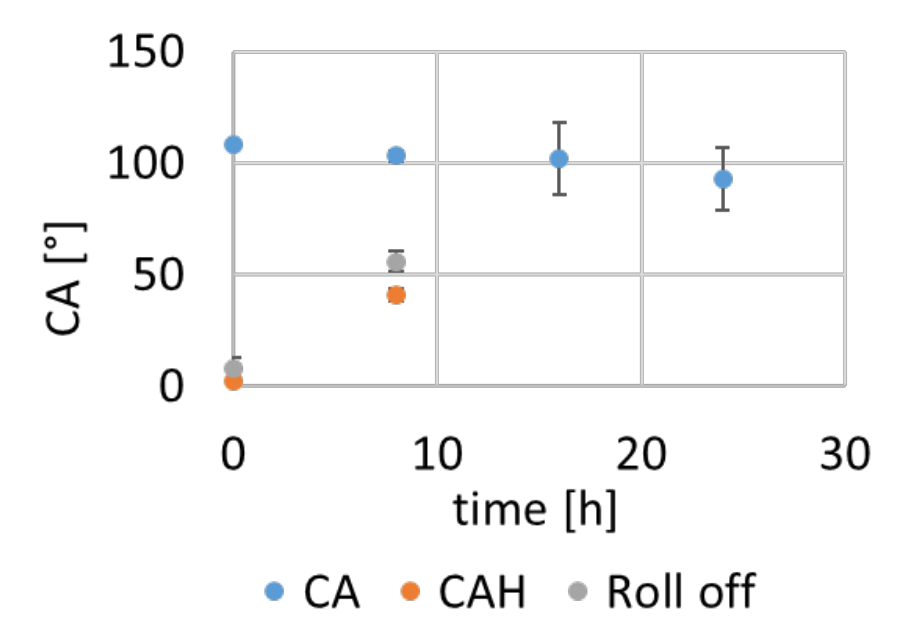}
		\caption{Static contact angle, contact angle hysteresis and roll off angle over time of operation}
		\label{statCAvst}
	\end{center}
\end{figure}

In Figure \ref{WettingComparison}, the surface wetting behavior of PDMS-coated and uncoated copper surfaces during condensation operation is displayed. 
\begin{figure}[!ht]
	\begin{center}
		\includegraphics[width=\textwidth]{fig/WettingComp2_compressed_compressed.pdf}
		\put(-390,8){$\SI{10}{mm}$}
		\caption{Surface wetting behavior during condensation.\\First row: uncoated copper substrate at wall subcoolings (left to right) $\SI{1.6}{K}$, $\SI{2.9}{K}$, $\SI{4.9}{K}$, $\SI{6.5}{K}$, $\SI{10.6}{K}$ \\Second row: PDMS-coated substrate on first day of condensation experiments at wall subcoolings (left to right) $\SI{2.3}{K}$, $\SI{2.8}{K}$, $\SI{3.7}{K}$, $\SI{4.9}{K}$, $\SI{6.2}{K}$
		\\Third row: PDMS-coated substrate on second day of condensation experiments at wall subcoolings (left to right) $\SI{2.1}{K}$, $\SI{3.6}{K}$, $\SI{5.0}{K}$, $\SI{6.4}{K}$, $\SI{7.5}{K}$
		\\Fourth row: PDMS-coated substrate on third day of condensation experiments at wall subcoolings (left to right) $\SI{2.4}{K}$, $\SI{3.6}{K}$, $\SI{5.2}{K}$, $\SI{6.4}{K}$, $\SI{8.0}{K}$}
		\label{WettingComparison}
	\end{center}
\end{figure}
In the first row, the surface wetting behavior on the uncoated surface is shown for various wall subcoolings. Initially, even on the uncoated copper surface, DWC is observed. The visible droplets show a strongly pinning wetting behavior and are irregular in shape. At increasing wall subcoolings, a transition regime is reached where some areas are covered with rivulets while other areas are covered by a continuous liquid film. At maximum wall subcooling, nearly all the surface is covered by the liquid film with an exception of a small area on right side, upper part of the surface. In the second row, the surface wetting behavior on the freshly PDMS-coated surface is shown for various wall subcoolings. With the increasing wall subcooling (from left to right), dropwise condensation is observed with an increasing number of visible drops on the surface. At the highest subcooling step in the end of the first measurement run, the drop sizes are increased compared to earlier recordings. The increased drop size indicates surface coating ageing, leading to reduced drop mobility and thus increased drop size before roll off. This trend of increased drop sizes before departure continues on the second day of condensation experiments, as illustrated in images displayed in the third row. In addition, the formation of filmwise condensation areas on the substrate starts at high values of subcooling.
On the third day of condensation experiments, displayed in the fourth row of Figure \ref{WettingComparison}, the filmwise condensation area grows in size until the large part of the substrate is covered by a liquid film at the last subcooling step.\\
In Figure \ref{SweepingComparison}, the droplet sweeping events on a $\SI{8}{mm} \times \SI{9.5}{mm}$ area in the substrate center on the first, the second and the third day of experiment (top to bottom row) at increasing values of wall subcooling (from left to right) are displayed. On all images, apart from the second image from the left in third row, the sweeping event is displayed, and the sweeping drop is highlighted. It can be seen that the average size of sessile condensate droplets at the moment of sweeping changes with increasing subcooling and with increasing surface age. On the first day of the experiment, the drops grow comparatively little before being swept by bigger drops rolling off. On the second day, a bigger population of larger drops can be observed on the surface at all subcooling steps. This trend continues to even larger drops on the third day of condensation operation. If the subcooling is fixed, the average size of the sessile condensate drops decreases with increasing frequency of the sweeping events. Therefore, the increasing size of the drops after the first day of the condensation experiment is consistent with the decreasing frequency of sweeping events with the surface ageing, which is discussed below. It has been also found that for all three days of condensation experiments, the number of downward moving droplets within the field of view per time interval increases with wall subcooling, leading to a higher removal rate of sessile condensate droplets.
\begin{figure}[!ht]
	\begin{center}
		\includegraphics[width=\textwidth]{fig/SweepingComp5_compressed_compressed.pdf}
		\put(-420,65){}
		\color{white}
		\put(-411,10){$\SI{2}{mm}$}
	    \color{black}
		\caption{Zoomed central rectangular region of substrate to detect droplet sweeping events on PDMS-coated substrate. The sweeping drops are highlighted with white dashed ovals.\\
		First row: First day of condensation experiments at wall subcoolings (from left to right) $\SI{2.3}{K}$, $\SI{2.8}{K}$, $\SI{3.7}{K}$, $\SI{4.9}{K}$, $\SI{6.2}{K}$
		\\Second row: PDMS-coated substrate on second day of condensation experiments at wall subcoolings (left to right) $\SI{2.1}{K}$, $\SI{3.6}{K}$, $\SI{5.0}{K}$, $\SI{6.4}{K}$, $\SI{7.5}{K}$
		\\Third row: PDMS-coated substrate on third day of condensation experiments at wall subcoolings (left to right) $\SI{2.4}{K}$, $\SI{3.6}{K}$, $\SI{5.2}{K}$, $\SI{6.4}{K}$, $\SI{8.0}{K}$}
		\label{SweepingComparison}
	\end{center}
\end{figure}
In Figure \ref{DropDiaDep}, the drop departure diameter on the PDMS-coated surface is plotted over the wall subcooling. At the beginning of the first day of condensation experiments, droplets start to accelerate downwards due to gravity at a diameter of $\SI{0.8}{mm}$ during the first three subcooling steps of $\SI{2.3}{K}$, $\SI{2.8}{K}$ and $\SI{3.7}{K}$. For the last two subcooling steps ($\SI{4.9}{K}$ and $\SI{6.2}{K}$), the drop departure diameter increases to $\SI{1.3}{mm}$. On the second day of condensation experiments, the drop departure diameter for the first two subcooling steps ($\SI{2.1}{K}$ and $\SI{3.6}{K}$) is increased to approximately $\SI{3.5}{mm}$ and later increases further to $\SI{4.2}{mm}$ for the wall subcooling steps of $\SI{5.0}{K}$ and $\SI{6.4}{K}$. For the second experiment day's last of the five subcooling steps, no roll off event happened during the 80 frame long recording. During the third day of the experiment, only during the second subcooling step, a roll off event happened during the recording time, showing an even further increased drop departure diameter of $\SI{5.6}{mm}$ at $\SI{3.6}{K}$ wall subcooling. In general, the longer the PDMS-coated substrate was used for condensation operation, the stronger is the droplets' mobility decreased as an increasingly higher single drop volume becomes necessary to start the acceleration process.\\
The reduction of mobility of drops can also be observed in Figure \ref{FreqSweeping} when comparing the frequency of sweeping events at different wall subcoolings for three subsequent days of condensation operation. During the first day of operation, the frequency of sweeping events in the substrate center is initially rising with wall subcooling, which is expected as higher wall subcooling causes more vapor to condense and subsequently to be drained. The increasing number of droplets leads to a more frequent sweeping of the surface. The sweeping is caused by draining droplets, which are accelerated downwards due to gravity. On their way down, they merge with sessile droplets and by that remove those from the condensation surface. The more droplets form on the surface and accelerate downwards with gravity, the higher the frequency of sweeping of surface area beneath the moving droplet. With a higher frequency of sweeping events for increased wall subcooling, the time for sessile droplets to grow and to increase in diameter reduces, which leads firstly to smaller droplets sitting on the surface and secondly to higher cleansing rate of nucleation sites. Both effects affect heat transfer positively \cite{Tripathy.2021, Weisensee.2017}.\\
For the last subcooling step of first day's experiment, a decrease in frequency of sweeping events is observed. This decrease can be attributed to surface ageing effects. This  suggestion is supported by increasing of condensate drops in size, as can be seen in Figure \ref{WettingComparison}, first picture from the right in the second row. The trend towards reduced frequencies of sweeping events continues after the first day of the condensation experiment. During second and third day of the condensation experiment the overall values of sweeping events is strongly decreased compared to the first day's values. On second and third day, the frequency of sweeping events increases with wall subcooling, but follows similar trends and values during both days, indicating no more changes in condensate-surface wetting behavior in the investigated central substrate region after the first day of condensation operation.\\
\\To further characterize the condensation operation induced changes in wetting behavior, contact angle measurements have been conducted on the PDMS-coated surface prior to and after each day of experiment. The results of measurements of the static contact angle as well as of the roll off angles are shown in Figure \ref{statCAvst}.
The initial values for static contact angle of $\SI{108}{\degree} \pm \SI{1}{\degree} $ and contact angle hysteresis of $\SI{2}{\degree} \pm \SI{1}{\degree} $ prior to the first experimental campaign match the values displayed in Figures \ref{1xARCA} and \ref{100xARCA}. In addition, a roll off angle of $\SI{8}{\degree} \pm \SI{4}{\degree} $ is determined for a $\SI{2}{\mu l}$ drop.\\
After 8 hours of operation, the static contact angle decreases to $\SI{103}{\degree} \pm \SI{2}{\degree} $, while a contact angle hysteresis of $\SI{41}{\degree} \pm \SI{3}{\degree} $ is determined. Here, a drop volume of $\SI{30}{\mu l}$ is used as no roll off was observed for smaller drops. A roll off angle of $\SI{56}{\degree} \pm \SI{5}{\degree} $ is determined.\\
While the trend for decreasing static contact angles with increasing hours of operation continues to  $\SI{92}{\degree} \pm \SI{14}{\degree} $ after 24 hours, contact angle hysteresis cannot be determined for higher experimental durations as no drop roll off was observed anymore for $\SI{90}{\degree} $ surface inclination and $\SI{30}{\mu l}$ drops. \\ 
In this study, three different measures have been used to quantify the surface wetting behavior during and after the condensation experiment and to characterize the surface ageing. The drop departure diameter is nearly independent of subcooling, but strongly changes from day to day of condensation operation. On the first day of condensation operation, the frequency of sweeping events initially strongly increases with increasing wall subcooling, reaches a maximum and decreases at the last subcooling step. On the second and third day, the increase of the frequency of the sweeping events with the subcooling is much weaker. Moreover, no substantial change of the behavior of the sweeping frequency from the second to the third day has been observed.
Contact angle measurements show a decrease in static CA and an increase in CA hysteresis, indicating a reduction in the PDMS coating's hydrophobicity. The results of all three characterization methods prove a strong change of the surface after the first day of the condensation experiment. The contact angle, contact angle hysteresis and drop departure diameter indicate a further change of the surface quality after the second day of the experiment. However, these changes did not affect the frequency of the sweeping events. The relationship between the reported characteristics will be investigated in the future.  \\
To determine the relevance of the different  characteristics of wetting behavior for condensation heat transfer, in the next section we compare the trends observed for wetting characteristics  with the trends of the condensation heat transfer parameters.
\subsection{Heat transfer}
\begin{figure}[htbp]
	\begin{minipage}[t]{0.49\textwidth}
		\includegraphics[width=0.99\textwidth]{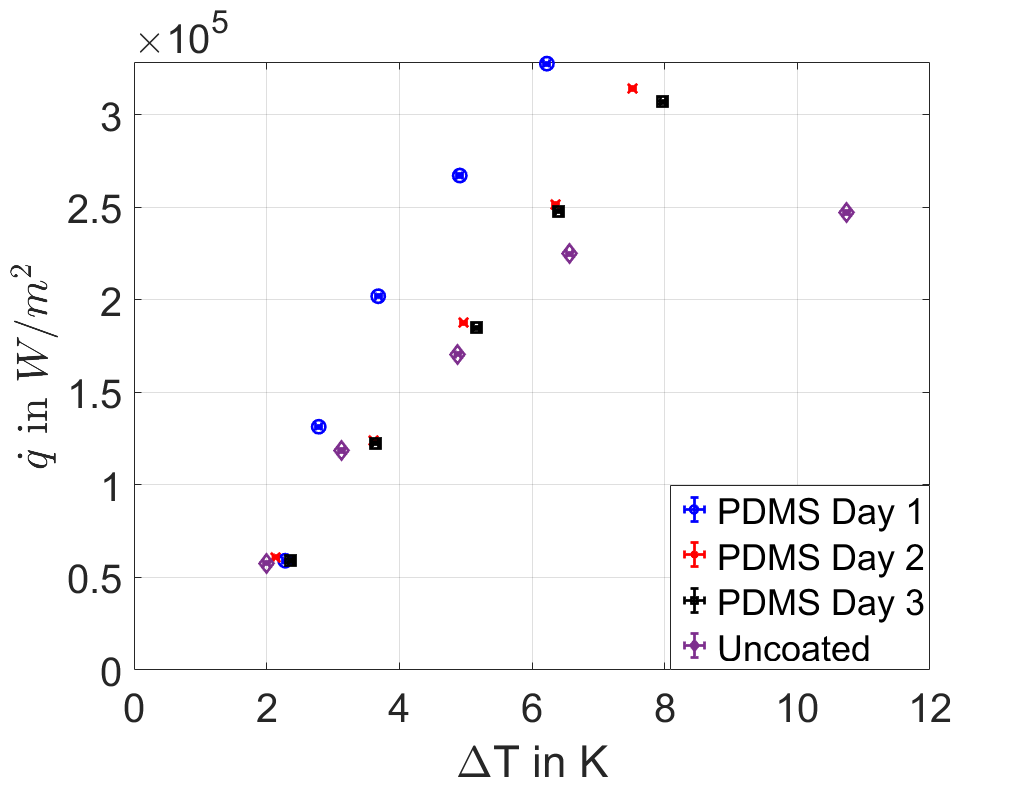}
		\caption{condensation heat flux over wall subcooling\\ for PDMS-coated and uncoated copper surfaces}
		\label{qvsdT}
	\end{minipage}
	\begin{minipage}[t]{0.49\textwidth}
		\includegraphics[width=0.99\textwidth]{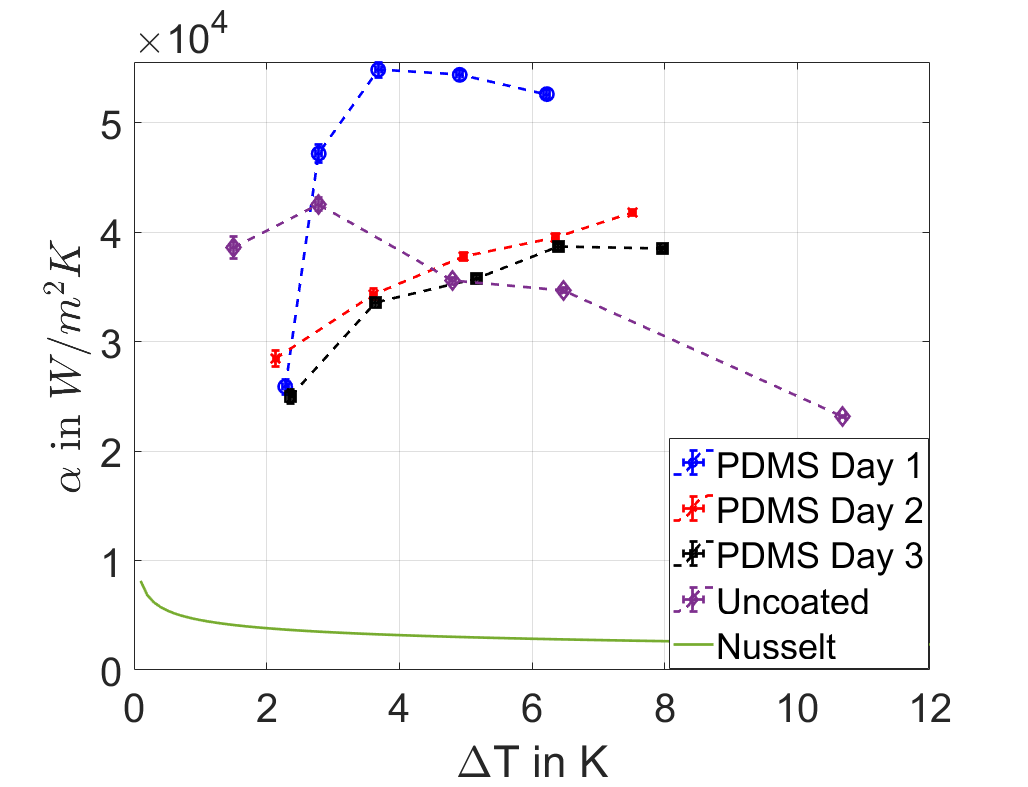}
		\caption{condensation heat transfer coefficient over wall\\ subcooling for PDMS-coated and uncoated copper surfaces}
		\label{alphavsdT}
	\end{minipage}
\end{figure}

\begin{figure}[!ht]
	\begin{center}
		\includegraphics[width=0.5\textwidth]{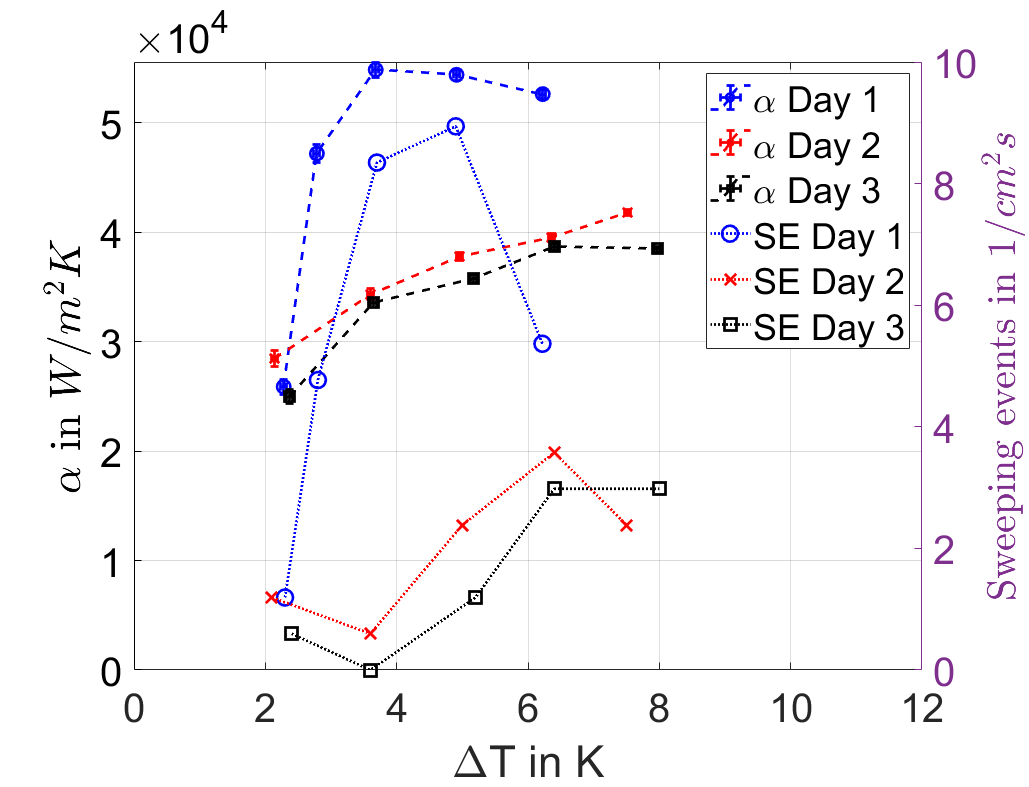}
		\caption{Heat transfer coefficient and frequency of sweeping events over wall subcooling}
		\label{alphaFrequencyDT}
	\end{center}
\end{figure}

In order to assess the PDMS-coating's impact on condensation heat transfer performance, the heat flux and heat transfer coefficient of both PDMS-coated and uncoated substrates are investigated. The uncoated substrates are fabricated using a process similar to that used for the PDMS-coated substrates, with the exception of the final PDMS-coating step.
In Figure \ref{qvsdT}, the heat flux over wall subcooling on both investigated surfaces is displayed. For the PDMS-coated surface, the heat flux for three consecutive days of experiments are displayed. On the polished and oxidized copper surface, represented by purple markers, an increase of heat flux over wall subcooling until a maximum of $\SI{247}{kW \per m^2}$ at $\SI{10.7}{K}$ was observed.
On the PDMS-coated copper substrate, increasing wall subcooling leads to a rise in heat flux on each measurement day. On the first day of experiment (blue markers), a significantly higher heat flux has been measured compared to the uncoated surface. On this day, a maximum heat flux of $\SI{327}{kW \per m^2}$ was reached at a wall subcooling of $\SI{6.2}{K}$. On the second day of experiment (red markers), the transferred heat fluxes are reduced compared to the first day of experiment. The maximum value of $\SI{314}{kW \per m^2}$ was reached at a wall subcooling of $\SI{7.5}{K}$. On the third day of operation (black markers) the heat no significant changes of the heat flux compared to the second day of operation were observed, and the maximum heat flux value of $\SI{307}{kW \per m^2}$ was reached at a wall subcooling of $\SI{8.0}{K}$.
Compared to an uncoated surface, the heat flux on a freshly coated substrate is increased by a factor of 1.6 at $\SI{5.0}{K}$ wall subcooling and a factor of 1.5 at $\SI{6.2}{K}$ wall subcooling. On second and third day of operation, the transferred heat flux is increased by a factor of only 1.1 at $\SI{5.0}{K}$ wall subcooling and at $\SI{6.2}{K}$ wall subcooling.
In Figure \ref{alphavsdT}, the heat transfer coefficient over wall subcooling on coated and uncoated surfaces is displayed together with the result of the Nusselt model for the filmwise condensation heat transfer. On the polished and oxidized copper surface, represented by purple markers, an initial rise in heat transfer coefficient from $\SI{39}{kW \per m^2K}$ at $\SI{1.5}{K}$ up to $\SI{43}{kW \per m^2K}$ at $\SI{2.8}{K}$ is followed by a decrease down to $\SI{23}{kW \per m^2K}$ at $\SI{10.7}{K}$.\\
On the freshly PDMS-coated copper substrate, represented by blue markers, an initial rise in heat transfer coefficient from $\SI{26}{kW \per m^2K}$ at $\SI{2.3}{K}$ up to $\SI{55}{kW \per m^2K}$ at $\SI{3.7}{K}$ is followed by a decrease down to $\SI{53}{kW \per m^2K}$ at $\SI{6.2}{K}$. On the second day of operation, represented by red markers, the heat transfer coefficient continuously rises from $\SI{28}{kW \per m^2K}$ at $\SI{2.1}{K}$ up to $\SI{42}{kW \per m^2K}$ at $\SI{7.5}{K}$. On the third day of operation, represented by black markers, the heat transfer coefficient results match those of second day of operation.\\
The heat transfer coefficient enhancement factors are equal to those presented on heat flux above. Namely, the enhancement factor of 1.5-1.6 has been observed for the freshly coated surface and an enhancement factor of 1.1 during the second and the third day of condensation operation.
Indicated in green, the analytical solution of heat transfer coefficient during filmwise condensation by Nusselt is displayed \cite{Nusselt.1916}. The heat transfer coefficient as computed by the Nusselt model weakly decreases with increasing of subcooling. The deviations of heat transfer coefficient measured on an uncoated copper surface from the predictions of Nusselt's model can be explained by analysing the wetting behavior during the condensation of water vapor on this surface. The model of Nusselt has been developed under an assumtion that the surface is fully covered by a condensate film, which drains down the surface in a laminar flow. These assumptions are obviously not valid for the  flow pattern on the uncoated copper surface, as can be seen in Figure \ref{WettingComparison}, top row.\\
When comparing the heat transfer performance of the PDMS-coated surface to the uncoated one, the initial similarity at low $\Delta T$ can by explained by comparing both surfaces' wetting behavior at low subcoolings. Both surfaces show DWC mode. Even though condensate droplets on the uncoated surface are comparatively large, the absence of hydrophobic coating's additional thermal resistance leads to an increased heat transfer and compensates the decrease of heat transfer due to increased drop size.\\
At higher values of the wall subcooling, the uncoated surface is increasingly covered by condensate, forming a liquid film separating large parts of the substrate from the vapor phase, resulting in a decreasing of the heat transfer coefficient with increasing of wall subcooling. Concerning the PDMS-covered surface, the continuously increasing heat transfer coefficient with the wall subcooling can be explained by considering the sustained DWC mode at all subcooling levels and only limited increasing of the surface coverage with condensate, except for the highest values of the wall subcooling at the third day of condensation experiment (see Figure \ref{WettingComparison}, bottom row). At higher wall subcoolings, the sweeping frequency is increased due to increased mass flow of condensate, which leads to less time needed for clearing drop nucleation sites and resetting the droplet nucleation process. The frequent clearing of the surface and resetting the nucleation process promotes the improvement of heat transfer \cite{Weisensee.2017,Sablowski2022ExperimentalAT}.\\
Indeed, one can observe that the measured heat transfer coefficient follows the trend of the frequency of the sweeping events (see Figure \ref{alphaFrequencyDT}). In particular, the dependence of the heat transfer coefficient and of the frequency of sweeping events on the wall subcooling is similar during the first day of experiment.  Followed by a strong initial increase, both the heat transfer coefficient and the sweeping frequency reach a maximum value at approximately the same value of subcooling and decrease  with the further increasing of subcooling. During the second and the third day of the condensation experiment, both the wetting behavior and the heat transfer are affected by the ageing of the surface coating. The heat transfer coefficient and frequency of sweeping events show no significant changes from the second to the third day of experiment. The heat transfer coefficient and the frequency of the sweeping events stay the same as in the first day for the lowest value of the subcooling. The ageing effect manifests itself in a much slower increasing of the heat transfer coefficient and of the frequency of the sweeping events with the subcooling. It can be seen that at the subcooling of around 6.4 K, the increasing trend of heat transfer coefficient and of the frequency of the sweeping events terminates.\\
The results presented in Figure \ref{alphaFrequencyDT} indicate that the heat transfer coefficient during the dropwise condensation correlates with the frequency of sweeping events which can be determined from optical data. The conditions and limitations of this interrelation will be determined in future experiments in an increased range of experimental parameters. Although the above correlation can be rationalized on the basis of principles of the condensation heat transfer, quantitative prediction of the interrelation requires a dedicated theoretical and numerical model.\\
The reasons of the ageing of the PDMS-coating, as well as the measures for the improvement of its robustness are presently under investigation. The simultaneous analysis of the heat transfer and wetting behavior characteristics, in particular the frequency of sweeping events, as a function of the subcooling and the duration of condensation operation, as proposed in the present work, is a valuable tool for monitoring the quality of the coating and its suitability for heat transfer applications. 
\section{Summary and conclusions}
\label{sec:Conclusions}
In this study, condensation of water on polydimethylsiloxane (PDMS)-coated copper surfaces is investigated. The presented surface fabrication method results in a contact angle value of $\SI{111}{\degree} \pm \SI{1}{\degree} $ during drop expansion and a stable value of $\SI{110}{\degree} \pm \SI{1}{\degree} $ during constant drop volume. The contact angle hysteresis of $\SI{2}{\degree}$ and its ultrathin coating thickness of $\SI{5}{nm}-\SI{10}{nm} $ indicate the PDMS-coating's suitability for dropwise condensation (DWC) applications. Repeated advancing and receding contact angle (ARCA) investigations support these findings, with consistent contact angle hysteresis of $\SI{2}{\degree}$ over 100 repetitions. These results align with previous studies on PDMS-coated glass substrates, reinforcing the effectiveness of PDMS coatings in promoting the dropwise condensation.\\
The PDMS-coated surface's wetting behavior during and after condensation experiments is investigated and compared to an uncoated surface. On the uncoated surface, DWC at low subcooling values is observed, followed by a transition regime with rivulets and liquid film coverage at higher subcoolings. On the PDMS-coated surface, initial DWC behavior is observed in a wide range of the subcooling values. Drop departure diameter remains constant, independent of subcooling, but increases on the second day and again on third day, limiting the droplet mobility. The frequency of sweeping events depends on the wall subcooling and is strongly affected by the surface ageing. On the first day of the experiment, higher subcoolings lead to more frequent sweeping events, until a maximum of the sweeping frequency is reached. After that the frequency decreases with a further increase of the wall subcooling. On the second day of the condensation experiments, the sweeping frequency is reduced. No substantial changes of the frequency have been observed on the third day in comparison with the first day. Static contact angle investigations and drop departure diameter reveal a decreasing surface wettability during the second and third day of operation. \\
Finally, the condensation heat transfer performances of PDMS-coated and uncoated substrates are compared with each other by evaluating the heat flux and heat transfer coefficient. On the first day of the condensation experiment, the heat flux on the PDMS-coated surface reaches a maximum of $\SI{327}{kW \per m^2}$ at a wall subcooling of $\SI{6.2}{K}$, 1.5-times that of the uncoated surface. On the second and third day of condensation experiments, the heat transfer enhancement factor reduces to a constant 1.1 compared to the uncoated surface. It has been found that the behavior of heat transfer coefficient correlates with the behavior of the frequency of the sweeping events on all three days of the condensation experiment.
\acknowledgements
The authors acknowledge the financial support by the German Research Foundation - Project number 265191195 - subproject C03.

\bibliographystyle{Bibliography_Style}

\bibliography{References}
\end{document}